\documentclass[aps,prb,twocolumn,showpacs,superscriptaddress,floatfix]{revtex4-1}

\pdfoutput=1

\usepackage{amsmath,amssymb}
\usepackage{graphicx}
\usepackage{color}
\usepackage{rotating}
\usepackage{bm}
\usepackage{setspace}
\usepackage{hyperref}
\usepackage{float}
\usepackage{soul}
\usepackage{color}
\usepackage{braket}

\hypersetup{
	colorlinks=true,
	urlcolor= blue,
	citecolor=blue,
	linkcolor= blue,
	bookmarks=true,
	bookmarksopen=false,
}	

\begin{document}
	
	\preprint{APS/123-QED}
	
	\title{Pressure dependence of the magic twist angle in graphene superlattices}

	\author{Stephen Carr}
	\affiliation{Department of Physics, Harvard University, Cambridge, Massachusetts 02138, USA.}
	\author{Shiang Fang}
		\affiliation{Department of Physics, Harvard University, Cambridge, Massachusetts 02138, USA.}
	\author{Pablo Jarillo-Herrero}
		\affiliation{Department of Physics, Massachusetts Institute of Technology, Cambridge, MA, USA.}
	\author{Efthimios Kaxiras}
		\affiliation{Department of Physics, Harvard University, Cambridge, Massachusetts 02138, USA.}
		
	\date{\today}
	
	\begin{abstract}

The recently demonstrated unconventional superconductivity~\cite{Cao2018sc}  
in twisted bilayer graphene (tBLG) opens the possibility for interesting applications 
of two-dimensional layers that involve correlated electron states.
Here we explore the possibility of modifying electronic correlations by the
application of uniaxial pressure on the weakly interacting layers, 
which results in increased interlayer coupling and a modification 
of the magic angle value and associated density of states.
Our findings are based on first-principles calculations that accurately describe
the height-dependent interlayer coupling through the combined use of
 Density Functional Theory and Maximally localized Wannier functions. 
 We obtain the relationship between twist angle and external pressure 
 for the magic angle flat bands of tBLG. 
 This may provide a convenient method to tune electron correlations by controlling the length scale of the superlattice. 
	
	\end{abstract}
	
	\maketitle
	
	
	Recent experimental results in twisted bilayer graphene (tBLG) have shown it to be an important system for understanding unconventional superconductivity \cite{Cao2018sc}, 
	and more generally correlated physics in two-dimensional (2D) 
materials \cite{Cao2018mott}.
This discovery comes after systematic development of experimental techniques which 
at present allow for twist angle control in stacked 2D heterostructures with a 
remarkable precision of $0.1^\circ$\cite{Cao2016,Kim2016,Kim2017,Yoo2018}.
In bilayer graphene, a relative twist between the layers by a ``magic'' angle
produces just the right amount of band hybridization to form flat bands near the Fermi level \cite{Li2009, Morell2010, Bistritzer2011, San-Jose2012, Wong2015, Brihuega2012, Luican2011}.
The flat bands have the majority of their electron density located 
at the AA-stacking regions of the moir\'e supercell.
As the Fermi velocity goes to zero, the scale of the electron kinetic energy 
falls below the scale of the two-particle Coulomb interaction, producing correlated 
behavior, although the precise mechanism for these effects is still a topic of active research.
Understanding the nature of the flat bands induced by the magic angle twist in tBLG 
is vital in studies of correlated electrons in 2D, and could lead to the  
discovery of other systems with similar behavior, generally referred to as ``twistronics'' \cite{Carr2017}.
We present here an \textit{ab-initio} study of how the interlayer electronic coupling in 
tBLG depends on external uniaxial pressure in the direction perpendicular to the layers, 
and how this pressure could act as a tuning parameter for correlated physics.
	
Manipulating superconductivity in tBLG by external pressure would 
follow the historic trend of using pressure to probe the nature of the superconducting $T_c$ \cite{Chu2005,Chu2009}.
The $T_c$ in conventional BCS superconductors usually decreases with pressure, but in unconventional superconductors pressure often increases $T_c$.
This is attributed to strong dependence of electronic correlation on external pressure, although the exact mechanism is not well understood and may vary between materials.
2D materials are particularly sensitive to pressure
along the direction perpendicular to the layers, as they 
are coupled through weak van der Waals interactions.
The mechanical effects of pressure on monolayer graphene have been documented through a variety of methods \cite{Proctor2009,Nicolle2011,Yankowitz2016}, 
and recently electronic transport measurements were performed on a graphene-hBN device under pressure \cite{Yankowitz2018}.
		
	\begin{figure}
		\includegraphics[width=1\linewidth]{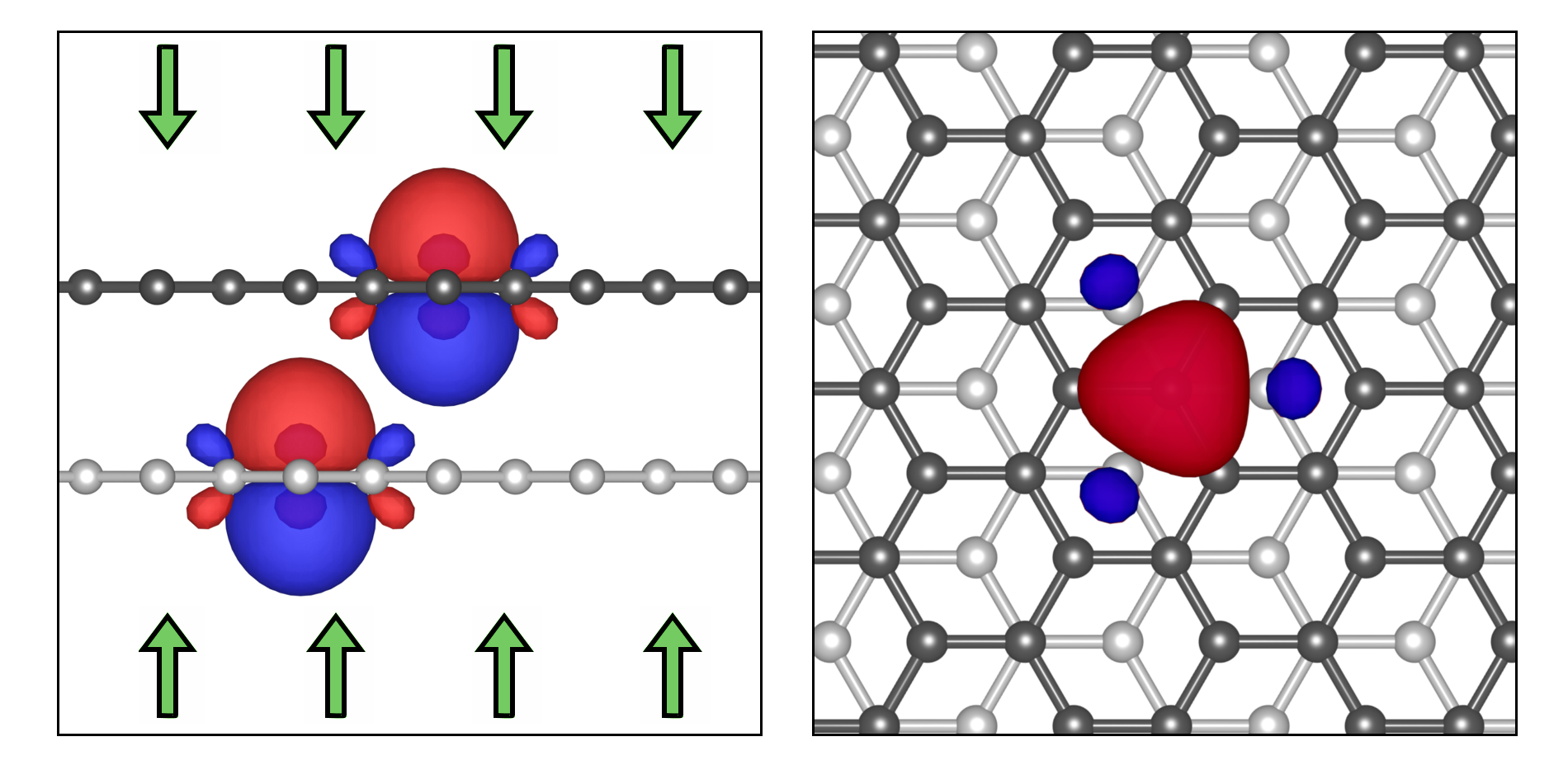}
		\caption{Isosurfaces of the localized Wannier functions in bilayer graphene, with the colors indicating different signs. (Left) Side view: Vertical compression of the bilayer mainly causes the orbitals to overlap more, thus increasing interlayer coupling while leaving in-plane couplings mostly unaffected.  
(Right) Top view: The triangular shape and nodes (indicated by the sign change) 
introduce angular dependence effects in the interlayer coupling, 
neglected in empirical tight-binding models for bilayer graphene based on $p_z$ orbitals.}
		\label{fig:wannier_bilayer}
	\end{figure}

In previous work we have derived \textit{ab-initio} tight binding hamiltonians for a range of 2D materials, including graphene, 
by using the maximally localized Wannier orbitals \cite{Marzari2012} to represent first-principles calculations based on density functional theory (DFT).
Our model for bilayer graphene identified a strong angular dependence of the interlayer coupling between the Wannier  orbitals \cite{Fang2016} (see Fig. \ref{fig:wannier_bilayer}).
To extend the model to compressed tBLG, we obtained the 
forces from DFT calculations \cite{Kresse1996}, including van der Waals corrections \cite{Peng2016}, 
from which we derive the relationship between external uniaxial pressure and interlayer distance in the bilayer 
as well as the pressure-dependent parameterization of the tight-binding hamiltonian.
Compression of the bilayer is given throughout the work in terms of $\epsilon = 1 - (d/d_0)$ where $d$ 
is the local interlayer distance and $d_0$ is the distance at zero external pressure ($d_0 = 3.35$ \AA \;from our calculations).

\begin{figure}
		\includegraphics[width=1\linewidth]{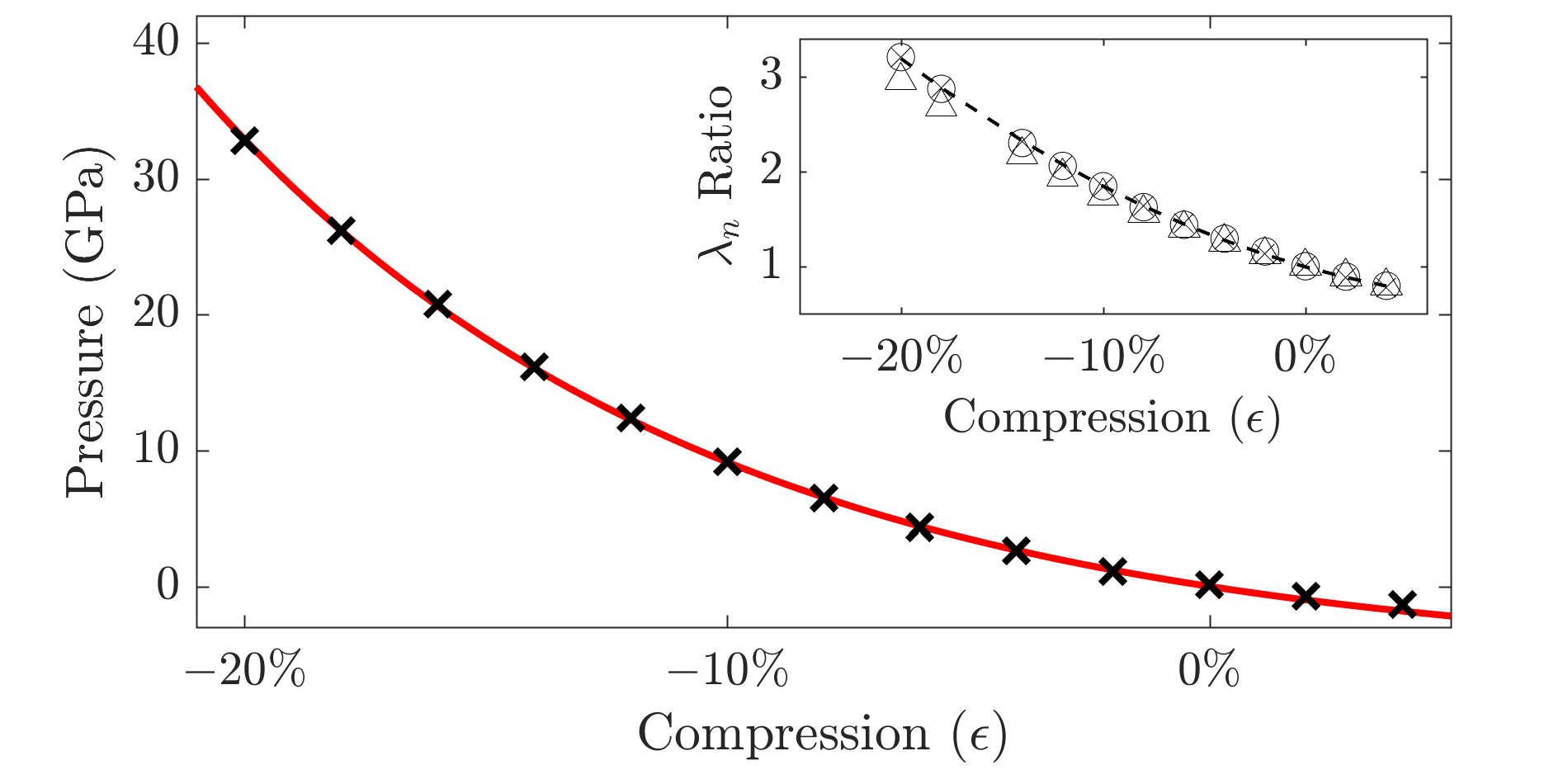}
		\caption{ 
		Calculated vertical external pressure as a function of interlayer distance between graphene layers (black crosses) with the the fit given in the text (red line).
		Inset: Compression dependence of the primary scaling parameters $\lambda_n$ of the interlayer coupling formula normalized by their values at $\epsilon = 0$. Values for $n = 0$, $3$, $6$ are shown with crosses, circles, and triangles respectively. The quadratic fit for $\lambda_0$ is given by the dashed line.}
		\label{fig:fitted_values}
	\end{figure}

The pressure from the DFT calculations is well fit by the functional form

	\begin{equation}
	P = A \left( e^{-B \epsilon} - 1 \right)
	\end{equation}

with $A = 5.73$ GPa and $B = 9.54$, as displayed in Fig. \ref{fig:fitted_values}.
We find that vertical compression of the bilayer 
has negligible effect on the in-plane tight-binding parameters, 
but significantly strengthens interlayer coupling.
The pressure dependence of the $10$ parameters of the interlayer coupling function\cite{Fang2016} are well described by a quadratic fit. 
The fit for the three scaling parameters of the interlayer coupling parameter $\lambda_n$ is given in the inset of Fig. \ref{fig:fitted_values}, where $n = 0,3,6$ corresponds to the three lowest channels of orbital angular momentum.
For more details we refer the reader to the supplementary material.

\begin{figure*}
		\includegraphics[width=1\linewidth]{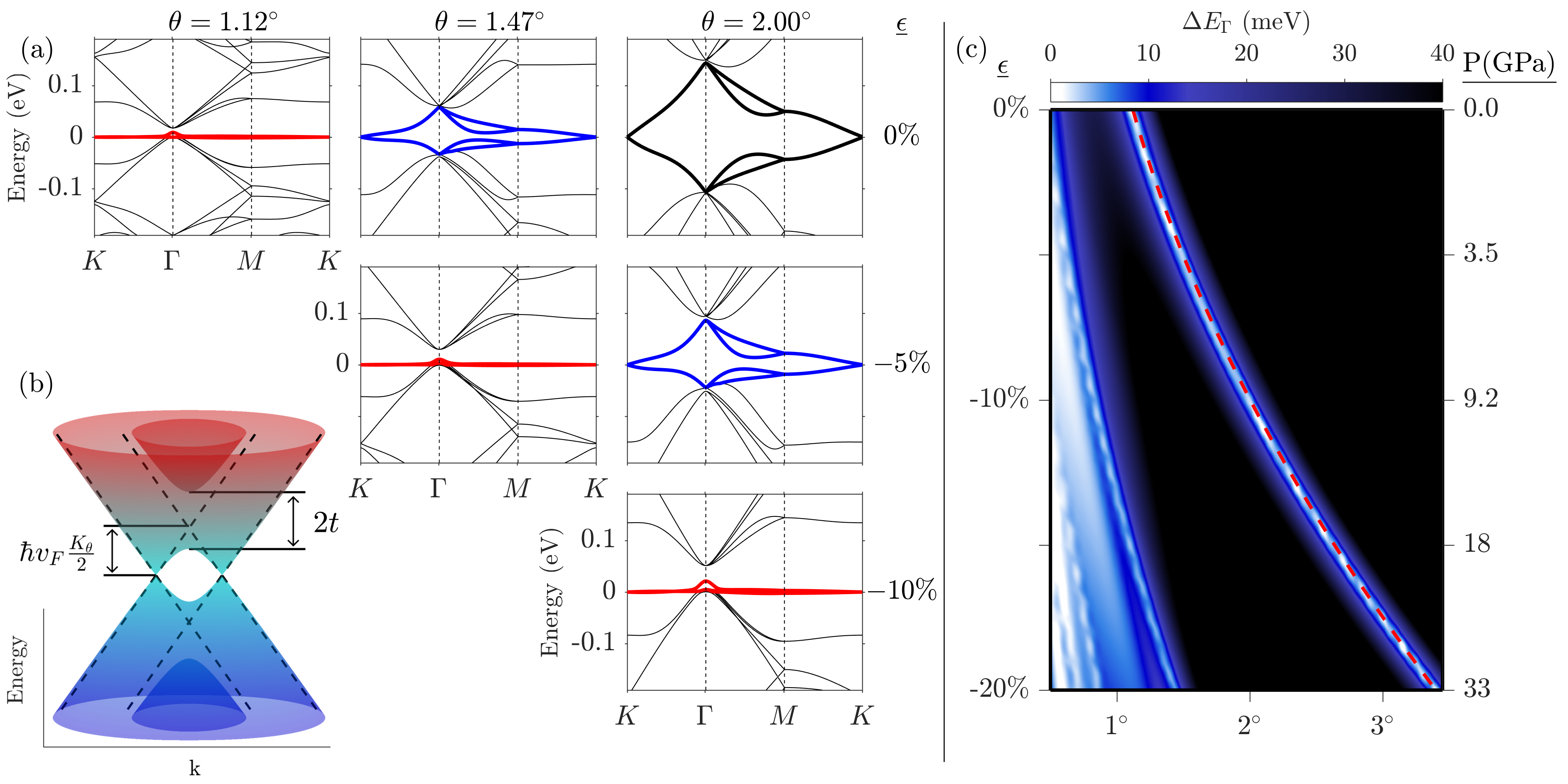}
		\caption{
\textbf{(a)} Band structures for twisted bilayer graphene under compression $\epsilon$ from the {\it ab initio} tight-binding model. 
The flat-band regime is achieved at $5\%$ compression for a twist angle of 
$1.47^\circ$, and at $10\%$ compression for a twist angle of $2.00^\circ$. 
\textbf{(b)} The two coupled Dirac cones, shifted in momentum space due to the twist, 
and with interlayer coupling strength $\lambda_0$, are shown schematically. 
\textbf{(c)} Critical values of the compression parameter $\epsilon$ as a function of twist angle. The bandwidth of the eight bands closest to the Fermi level at the $\Gamma$ point, $\Delta E_\Gamma$, is shown in color with white representing the small bandwidth of the flat bands. 
The dashed red line gives the expected value of the compression to cause flat bands
(see text for details).}
		\label{fig:magic_angle_bands}
	\end{figure*}

	In Fig. \ref{fig:magic_angle_bands}(a) we show the low energy electronic structure of tBLG 
at three different twist angles and compressions, calculated with a supercell tight-binding model.
	The magic angle can be thought of as a resonance of the bilayer hybridization, where the twist angle acts as a ``knob'' tuning the electronic structure \cite{Carr2017}.
	As the compression increases and the layers come closer to one another 
the effective interlayer coupling strength increases, causing stronger electronic hybridization between them.
	In particular, while the zero-pressure magic angle occurs at approximately 
$1.1^\circ$, under $10\%$ compression (9.2 GPa)
the magic angle is approximately $2.0^\circ$.
	Our calculations did not show significant reconstruction of the graphene bilayer under pressure even up to $30$ GPa,
	but there may be a phase transition of the encapsulating hBN substrate around $9$ GPa \cite{Yankowitz2018}.

	A heuristic argument for tracking the magic angle as
a function of compression or twist angle can be 
constructed from the perspective of coupled states in momentum space,
as shown in Fig. \ref{fig:magic_angle_bands}(b).
Without interlayer coupling, the low-energy band structure of a bilayer 
resembles two Dirac cones separated in momentum space by 
$K_\theta \approx G \theta$, where $G$ is the characteristic length of the reciprocal-cell 
lattice vectors of the monolayer. 
For simplicity, we avoid the complexities of scattering in momentum space 
that the twist angle introduces, and focus on the Bloch states 
which are exactly halfway between the $K$ and $K'$ points of the supercell (the $M$ point).
Before hybridization, 
each layer contributes two Bloch states with energies $\pm (K_\theta/2) \hbar v_F $, where $v_F$ is the Fermi velocity of a graphene monolayer.
We expect eigenvalues near $0$ when the interlayer coupling terms are equal in magnitude to the Bloch state energies.
A derivation including nearest-neighbor momentum scattering can give a more precise relationship for these terms \cite{Bistritzer2011}, but for our argument this is not necessary as we will only be interested in the relative scaling of inter and intralayer energies. 
We then assume this interlayer coupling strength $t$ has at most quadratic dependence on compression,

	\begin{equation}
	t(\epsilon) = t_2 \epsilon^2 - t_1 \epsilon + t_0 \propto \hbar v_F \frac{K_\theta}{2}.
	\end{equation}

Taking into account that $K_\theta \propto \theta$ and that there is a magic angle at zero compression ($\epsilon = 0$) of
approximately $\theta_0 = 1.12^\circ$, we can make the substitution 
$\hbar v_F (K_\theta/2) \to \theta 
( t_0/\theta_0)$ to obtain:

	\begin{equation}
	t_2 \epsilon^2 - t_1 \epsilon + t_0(1 - \theta/\theta_0) = 0
	\end{equation}
	
which gives the critical value, $\theta_c(\epsilon)$, of the magic
angle as a function of compression $\epsilon$:
	\begin{equation}
	\theta_c(\epsilon) = \theta_0 \left[ (t_2/t_0) \epsilon^2 - (t_1/t_0) \epsilon + 1 \right].
	\end{equation}
	
From this expression, we deduce that for experimentally accessible pressures
 any angle in the range [$1.1^\circ, 3.0^\circ$] can serve as the magic twist angle 
that leads to correlated behavior, by adjusting the pressure.
	To confirm this claim, we use an \textit{ab-initio} $k \cdot p$ model, as described in a previous work \cite{Massatt2018}, to sample the electronic bandstructure of tBLG under compression and with a twist angle in the estimated range. 
	To quantify the ``flatness'' of the band-structure we compute 
the bandwidth of the two bands closest to the Fermi level at the $\Gamma$ point
(Brillouin zone center). 
	As seen in Fig. \ref{fig:magic_angle_bands}(a), the bandwidth of the low-energy states 
at this point is on the order of a few meV, which 
gives a reliable indication of how flat the bands are. 
In Fig. \ref{fig:magic_angle_bands}(c) 
	we show the bandwidth at the $\Gamma$ point, referred to here as 
$\Delta E_\Gamma$, as a function of twist angle and compression. 
	The most prominent feature (white line) corresponds to the first magic angle value.
The lines at smaller angles correspond 
to higher-order magic angles.
The relationship derived from our heuristic argument agrees extremely well with numerical results if we use the values of the leading scaling parameter of the interlayer coupling, $\lambda_0$, and reduce the linear and quadratic terms by 8\%.
Additional parameters for the interlayer coupling beyond $\lambda_0$ are required 
to describe the pressure effects accurately, such as 
the angular distribution and range of the coupling. 
Their inclusion in the model would affect the interlayer coupling 
in $k \cdot p$ theory, leading to this small correction.
The corrected values are $t_{[0,1,2]} = [0.310, 1.731, 7.122]$ eV.
		
	As a final ingredient to enhance the reliability of the theoretical model, we use the distance dependent interlayer coupling to examine the effects of atomic relaxation in tBLG systems at $0$ pressure.
	The uncompressed bilayer exhibits significant relaxation at a twist angle of approximately 
$1^\circ$ \cite{Dai2016, Zhang2018, Yoo2018}.
This causes important changes to the low-energy bandstructure \cite{Nguyen2017}.
	Just as compression enhances electronic coupling between the layers, it also enhances atomistic coupling. 
	At large compression, significant relaxation is likely to occur at larger angles, including those that lead to flat bands under external pressure.
	 
\begin{figure}%
		\includegraphics[width=1\linewidth]{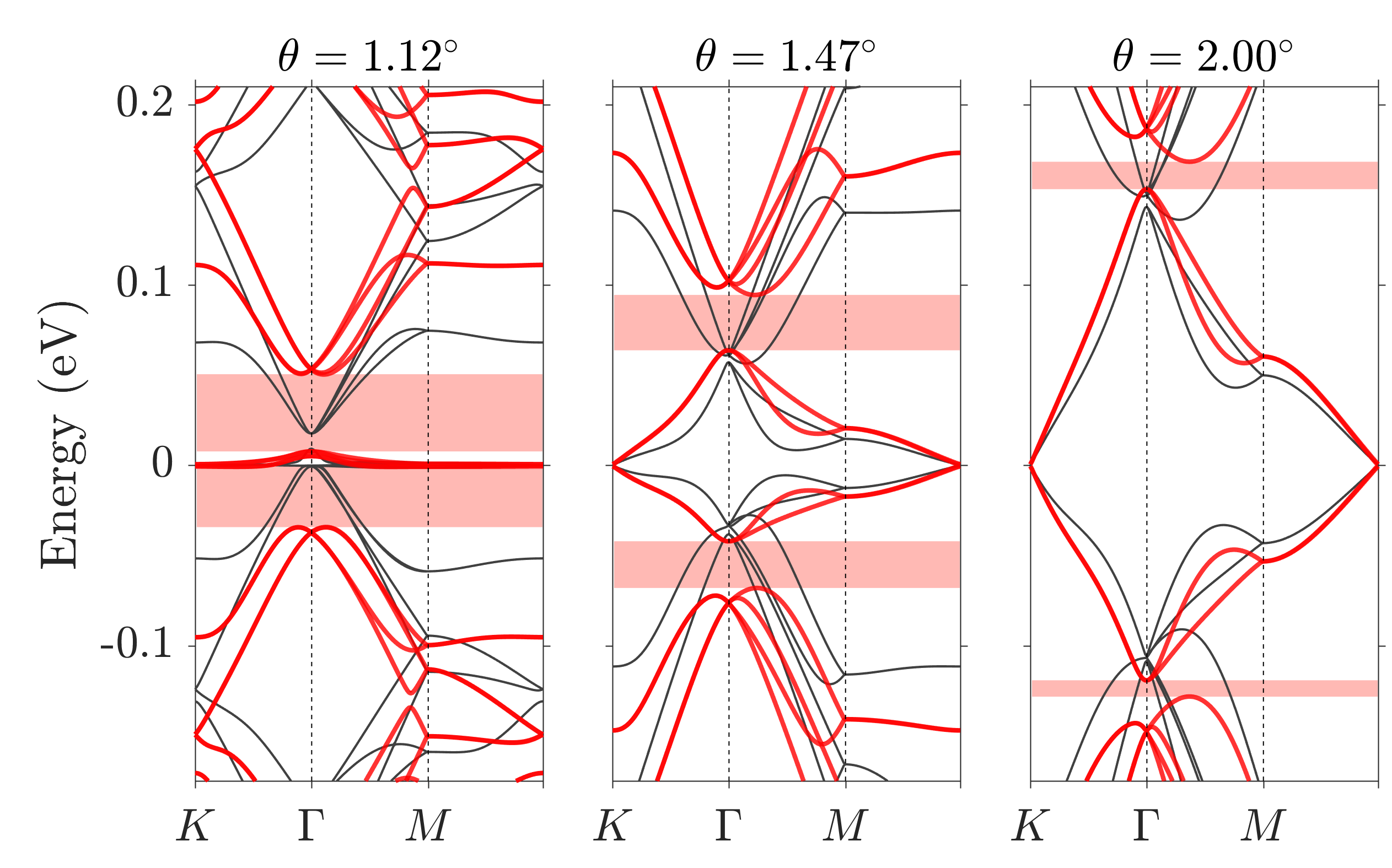}
		\caption{Band structures of uncompressed twisted bilayer graphene with and without relaxation of the atoms. The black lines are bands for the unrelaxed system and the red lines for the relaxed system. The single particle gaps in the relaxed system are highlighted in pink.}
		\label{fig:relaxed_bands}
	\end{figure}

	 In Fig. \ref{fig:relaxed_bands} we present our \textit{ab-initio} tight-binding 
bandstructure results for tBLG with and without relaxation.
The relaxation is taken into account by using a continuum model that uses only DFT values from generalized stacking fault calculations \cite{Zhou2015} adapted for twisted systems \cite{Carr2018}.
This model relaxes both the in-plane and the out-of-plane positions of the atoms and updates the interlayer coupling accordingly.
	The structure of the flat bands and the size of the single-particle gaps change with this correction, 
indicating that experimental study of correlation effects can depend sensitively
on the sample's environment and substrate effects.
	We find that relaxation increases the dispersion of the low energy bands
and increases the gaps on both sides to roughly $50$ meV near the magic angle, which is in good agreement with experiment \cite{Cao2016,Cao2018mott}.
	At larger angles the relaxation is less extreme, and the gap size decreases with increasing angle.
	Near $2^\circ$ the gaps are almost completely gone as the unrelaxed and relaxed bilayer geometry become similar.
	Quantifying the degree of relaxation in experimental devices will be an important ingredient for understanding the low energy electronic structure, and thus the superconductivity phenomenon, in graphene.

In conclusion, we have studied the behavior of flat bands induced by magic angle twist
in bilayer graphene as a function of external pressure.
The height dependent coupling allows for accurate band structure calculation for relaxed systems, showing that relaxation can play an important role in interpreting the low energy states of twisted bilayer graphene.
We demonstrated how the pressure may be used to produce correlated behavior, 
identified by the presence of flat bands at twist angles that increase with increasing pressure.
The larger twist angles lead to a moir\'{e} cell of smaller size, which is likely beneficial to the coupling strength and may enhance correlated electron behavior, including the superconducting $T_c$.
In the absence of clear understanding of the superconducting state it is impossible to provide 
{\em quantitative} predictions for these effects. 
Inverting the argument, we propose that systematic experimental study 
of correlated behavior as a function of pressure could shed light on 
the nature of unconventional superconductivity in tBLG and related systems. 

\begin{acknowledgements}
We acknowledge Y. Cao and V. Fatemi for helpful discussions. Computations were performed on the Odyssey cluster supported by the FAS Division of Science, Research Computing Group at Harvard University.
This work was supported by the ARO MURI Award No. W911NF-14-0247, 
the STC Center for Integrated Quantum Materials funded by NSF Grant No. DMR-1231319, 
and the Gordon and Betty Moore Foundation’s EPiQS Initiative through Grant  GBMF4541.
\end{acknowledgements}	
	
\bibliographystyle{apsrev4-1}
\bibliography{compressed_tblg}	

\clearpage

\section{Supplementary Materials}	

	In previous works, we reported the development of \textit{ab-initio} tight-binding models for 2D systems using a combined Density Functional Theory with Maximally Localized Wannier Functions approach (DFT+MLWF) \cite{Fang2015, Fang2016}.
	This process is done by first using a conventional DFT code (we use VASP \cite{Kresse1996}) to find the electronic-ground state Kohn-Sham wavefunctions in a standard plane-wave basis, $\psi_i(k)$, and then transforming these into a localized real-space basis, $\phi_i(r)$.
	After the transformation, the DFT hamiltonian can be used to compute the energy overlap matrix elements $\braket{\phi_i|H|\phi_j} = t_{ij}$, which give the hopping parameters of a tight-binding model.
	A 4-band tight-binding model of bilayer graphene was determined, consisting of intralayer hopping energies and a functional form for interlayer coupling.
	This interlayer coupling function included angular dependence due to the triangular warping of the $p_z$ type orbitals of carbon in the simplified graphene model.
	It is given as a sum of three terms, representing the different angular momenta of the wavefunctions:
	
	\begin{equation}
	\begin{split}
	t(\textbf{r}) = V_0(r) + & V_3(r) [\cos(3 \theta_{12}) + \cos(3 \theta_{21})] \\
	  + & V_6(r) [\cos(6 \theta_{12}) + \cos(6 \theta_{21}) ]
	\end{split}
	\end{equation}
	
	with the radial functions given by
	
	\begin{equation}
	\begin{split}
	V_0(r) &= \lambda_0 e^{- \xi_0 (\bar{r})^2} \cos(\kappa_0 \bar{r})		\\
	V_3(r) &= \lambda_3 \bar{r}^2 e^{-\xi_3 (\bar{r} - x_3)^2}				\\
	V_6(r) &= \lambda_6 e^{-\xi_6 (\bar{r} - x_6)^2} \sin(\kappa_6 \bar{r})    
	\end{split}
	\end{equation}
	
	where $\bar{r}$ is $r/2.46$\AA \;, the in-plane radius reduced by the in-plane lattice parameter. 
	This form takes into account both the in-plane radius and the relative angles between the displacement vector and the monolayer lattices.

	We find that when changing the interlayer distance of the bilayer by +4\% to -20\%, this form of interlayer coupling still agrees well with \textit{ab-initio} results and that the intra-layer couplings have negligible variation.
	The compression dependence of the electronic model can therefore be completely described by understanding how the 10 parameters in $t(r)$ change as the interlayer distance is modified.
	We use the same DFT+MLWF approach to model the interlayer coupling without allowing for structural relaxation of the individual monolayers.
	Although in a free-floating bilayer system the lattice parameters are likely to change under compression, most experimental studies create these devices by encapsulating them in insulating substrates, usually hBN.
	This encapsulation technique may change the in-plane lattice parameter as well, and so for simplicity we have ignored these effects.
	To compare the compression parameter $\epsilon$ to an experimental pressure of an encapsulated system, we have also calculated the external pressure of a bulk system consisting of three 2D layers: AB bilayer graphene separated by a single layer of hBN. 
	We approximate the lattice-paramter of hBN as equal to that of graphene, 2.46 \AA , meaning the system consists of only six atoms, 2 from each layer.
	Energies are computed in the VASP DFT software package with the van der Waals DFT method SCAN+rVV10 of Peng et al. \cite{Peng2016}, a k-mesh of $21 \times 21 \times 1$, and an energy cutoff of 500 eV.
	We simulate pressure by changing the height of the periodic cell and allow the graphene atoms to fully relax, but fix the locations of the hBN atoms.
	No significant restructuring of the graphene bilayer due to compression was observed.
	Running a calculation for multiple $\epsilon$ values in our sampling range yields a good fit for the external vertical pressure:
	
	\begin{equation}
	P = A \left( e^{-B \epsilon} - 1 \right)
	\end{equation}

	with $A = 5.73$ GPa and $B = 9.54$.
	For example, this gives $0$ GPa, $9.15$ GPa, and $32.89$ GPa at $\epsilon$ values of $0\%$, $-10\%$, and $-20\%$ respectively.
	
	From the DFT+MLWF calculations, we fit a quadratic model to each parameter in the interlayer coupling formula,
	
	\begin{equation}
	y_i(\epsilon) = c_i^{(0)} + c_i^{(1)} \epsilon + c_i^{(2)} \epsilon^2 
	\label{eq:quad_model}
	\end{equation}

	where $y_i$, $i = 1, \dots, 10$ represents one of the $10$ parameters of the model.
	We report the results of this fitting for each of the 10 parameters in Table \ref{tab:params}.
	
	\begin{table}
	\begin{center}
	\begin{tabular}{|l||r|r|r|}
	\hline
		$i$ $(y_i)$			& 	$c_i^{(0)}$	& 	$c_i^{(1)}$ 	& 	$c_i^{(2)}$ \\
		\hline
		$1$ $(\lambda_0)$ 	&  0.310 & -1.882 & 7.741 \\
		$2$ $(\xi_0)$ 		&  1.750 & -1.618 & 1.848 \\
		$3$ $(\kappa_0)$	&  1.990 &  1.007 & 2.427 \\
		\hline
		$4$ $(\lambda_3)$	& -0.068 &  0.399 & -1.739 \\
		$5$ $(\xi_3)$ 		&  3.286 & -0.914 &  12.011 \\
		$6$ $(x_3)$ 		&  0.500 &  0.322 &  0.908 \\
		\hline
		$7$ $(\lambda_6)$ 	& -0.008 &  0.046 & -0.183 \\
		$8$ $(\xi_6)$ 		&  2.272 & -0.721 & -4.414 \\
		$9$ $(x_6)$ 		&  1.217 &  0.027 & -0.658 \\
		$10$ $(\kappa_6)$ 	&  1.562 & -0.371 & -0.134 \\
		\hline
	\end{tabular}
	\caption{Fitted compression dependence for the 10 parameters of the interlayer coupling model. All parameters are given in units of eV and take the form given in Eq. \ref{eq:quad_model}.}
	\label{tab:params}
	\end{center}
	\end{table}		
	
	The three $\lambda_n$ parameters ($n = 0,3,6$) have the strongest dependence on $\epsilon$, while every other parameter is only weakly dependent. 
	This makes sense, as the $\lambda_n$'s set the overall strength of the electronic coupling between the layers and should increase quickly as the layers are forced closer together. 
	The other parameters encode angular and radial-centering information of the interlayer coupling, and are thus less affected by compression. 
	
\end{document}